\documentclass[preprint,12pt,authoryear]{elsarticle}

\usepackage{lmodern,sansmath}
\usepackage{textcomp}
\usepackage{microtype}
\usepackage{amsmath}
\usepackage{amssymb}
\usepackage{booktabs}
\usepackage{graphicx}
\usepackage{xcolor}

\def\pp#1#2{\frac{\partial #1}{\partial #2}}
\renewcommand{\vec}[1]{\boldsymbol{#1}}

\newcommand{\avg}[1]{\langle{#1}\rangle}

\journal{IJHFF}

\begin{document}

\begin{frontmatter}



\title{\textbf{ High-order structure functions for passive scalar {\color{black} fed by a} 
		mean gradient}} 


\author{M.\ Gauding, L.\ Danaila and E.\ Varea}

\address{CORIA  UMR 6614, Universit{\'e} de Rouen, 76801 Saint Etienne du Rouvray, France}

\begin{abstract}
  Transport equations for even-order structure functions are written
  for a passive scalar mixing fed by a mean scalar gradient, with a
  Schmidt number $\mathit{Sc}=1$.  Direct numerical simulations (DNS),
  in a range of Reynolds numbers $R_\lambda \in [88,529]$ are used to
  assess the validity of these equations, for the particular cases of
  second-and fourth-order moments.  The involved terms pertain
  to molecular diffusion, transport, production, and
  dissipative-fluxes.
  The latter term, present at all scales, is equal to:  i) the mean scalar variance dissipation rate,  $\langle \chi \rangle$,  for the second-order moments transport equation;  ii) non-linear correlations between $\chi$ and second-order moments of the scalar increment, for the fourth-order moments transport equation. \\
  The  equations are further analysed to show that the
  similarity scales (i.e., variables which allow for perfect collapse
  of the normalised terms in the equations) are, for second-order
  moments, fully consistent with Kolmogorov-Obukhov-Corrsin (KOC)
  theory. However, for higher-order moments, adequate similarity
  scales are built from $\langle \chi^n \rangle$.  The similarity is
  tenable for the dissipative range, and the beginning of the scaling
  range.
\end{abstract}


\end{frontmatter}


\section{Introduction}

%The local and non-local phenomena which are inherent to turbulent
%flows can be analyzed through the moments of the scalar increment, the
%so-called structure functions, defined by
%\begin{equation}
%  S^{(n)}(\vec x,\vec r) = \avg{(\phi(\vec x+ \vec r) - \phi(\vec x))^n} \,,
%\end{equation}
%where $\vec r$ is the separation vector between the two points and the
%angular brackets denote an ensemble-average. In statistically
%homogeneous turbulence the ensemble-average of scalar fluctuations is
%independent of the position $\vec x$.  The mean scalar gradient
%$\Gamma$ (aligned with say, $x_2$ direction) injects energy which maintains the fluctuating field 
%in a statistically stationary state. However,  $\Gamma$  imposes an
%anisotropy which persists even at the
%smallest scales, and as a result, structure functions depend not only
%on the magnitude $r$ of the separation vector but also on its
%orientation. \cite{gotoh2011} showed that statistical quantities of
%the scalar fluctuations reveal an axisymmetry around the $x_2$-axis
%and reflectional invariance to the plane perpendicular to the mean
%gradient.

Fully developed turbulence is characterized by a large range of length
scales, varying from the so-called integral length scale $l_t$, at
which large velocity fluctuations occur on average, down to the
smallest scale, the so-called Kolmogorov or dissipation scale $\eta$,
at which turbulent fluctuations are dissipated. Until now, most
understanding of turbulent flows has been gained from Kolmogorov's
scaling theory \cite{kolmogorov1941,kolmogorov1941b}, which was later
extended by \citet{obukhov1949} and \citet{corrsin1951} to passive
scalars advected by a turbulent velocity field.  The
Kolmogorov-Obukhov-Corrsin (hereafter, KOC) theory postulates that, under the condition
of sufficiently high Reynolds numbers, the small scales of the flow
decouple from the large scales. The understanding is that there is a
steady cascade from the large scales to the smallest scales, where the
energy transfer rate is equal to the mean energy dissipation rate
$\avg{\varepsilon}$. Kolmogorov hypothesized that the small scales
should depend only on two parameters, namely the viscosity $\nu$ and
the mean energy dissipation $\avg{\varepsilon}$. Because only two
quantities with different physical units are involved, this was viewed
as a claim of universality.  If the notion that the small-scale motion
is universal was strictly valid, then there would be realistic hope
for a statistical theory of turbulence. % The small scales of the scalar field
% additionally depend on the molecular diffusivity $D$ and the mean
% scalar dissipation $\avg{\chi}$.

The local and non-local phenomena which are inherent to turbulent
flows can be analyzed by the moments of the scalar increment
$\Delta\phi$, the so-called structure functions, defined by
\begin{equation}
  S_{\color{black} p} (\vec x,\vec r) = \avg{(\phi(\vec x+ \vec r) - \phi(\vec x))^{\color{black} p}} = \avg{(\Delta\phi)^{{\color{black} p}}} \,,
\end{equation}
where $\vec r$ is the separation vector between the two points and the
angular brackets denote an ensemble-average. In statistically
homogeneous turbulence,  structure functions are independent of the
position $\vec x$. \citet{Yaglom_1949} presented an exact transport
equation for the second order scalar structure function in homogeneous
isotropic turbulence, where the separation distance $r$ (the modulus of  vector $\vec r$)  is the
independent variable.\, i.e.
\begin{equation}
  \label{eq:yaglom}
  \pp{}{t}\avg{(\Delta\phi)^2} +
  \pp{}{r_i}\avg{(\Delta u_i)(\Delta \phi)^2} = 2 D \pp{^2}{r_i^2}
  \avg{(\Delta \phi)^2} - 2 \avg{\chi} \,,
\end{equation}
where $\Delta u_i=u_i(\vec x + \vec r) - u_i(\vec x)$ denotes the
velocity increment and $D$ the molecular diffusivity. Equation
\eqref{eq:yaglom} uses Einstein's summation convention, which implies
summation over indices appearing twice. The mean scalar dissipation
is defined as,
\begin{equation}
\avg{\chi} = 2D \left\langle{\left( \pp{\phi}{x_i} \right)^2}\right\rangle \,.
\end{equation}
Provided that the Reynolds number is high enough,
Eq.~\eqref{eq:yaglom} has two separate analytical solutions. One in
the dissipative range for $r\to 0$, where the diffusive term and the
scalar dissipation balance, i.e.
\begin{equation}
  \avg{(\Delta\phi)^2} = \frac{\avg{\chi}}{6D}\, r^2 \,,
\end{equation}
and one for the inertial range for $\eta \ll r \ll l_t$, where the
transport term balances the scalar dissipation, i.e.
\begin{equation}
  \avg{(\Delta u_L)(\Delta \phi)^2} = -\frac{2}{3} \avg{\chi} r \,,
\end{equation}
with $\Delta u_L$ being the longitudinal velocity increment in the
direction of $\vec r$. These two results are of high
significance. They are both exact and were derived from
first-principles only under the assumptions of (local) isotropy and
(local) homogeneity.

% The main aim of the present paper is to examine the Reynolds number
% dependence of a generalized Yaglom equation for higher order structure
% functions and wsze"'rdcf xfinite Reynolds numbers.

The paper is devoted to the analysis of $S_{\color{black} p}$ for ${\color{black} p}=2$ and ${\color{black} p}=4$, in
the context of transport equations obtained from the first principles,
by considering a restricted number of additional hypotheses, such as  self-preservation,
cf.~\citet{Danaila_Mydlarski_PRE01}.  Section \ref{section_DNS}
describes the main characteristics of the performed direct numerical
simulations (DNS), on which the analysis is based. In
Section ~\ref{section_theory} we present the theory of generalized scalar
scale-by-scale structure functions for any even order.  Next, we
develop in Section ~\ref{sec:similarity} similarity scales based on the
scale-by-scale budget equations for the second and fourth order
moments. The similarity scales are then justified by using data from
DNS with different Taylor length-scale based Reynolds numbers between
88 and 529. Concluding remarks are given in
Section ~\ref{sec:conclusions}.

\section{Direct numerical simulations} \label{section_DNS}
We study a passive scalar advected by a statistically homogeneous
isotropic and incompressible turbulent velocity field. In the present
study,  we consider a passive scalar with unity Schmidt number
$\mathit{Sc}=\nu/D$, so that the kinematic viscosity $\nu$ equals the
molecular diffusivity $D$. A uniform mean gradient $\Gamma$ is imposed
on the scalar field in $x_2$-direction. The mean gradient injects
continuously energy into the scalar field and keeps statistics in a statistically steady state.  The instantaneous scalar
field can be decomposed in a mean part $\Gamma x_2$ and a scalar
fluctuation $\phi$, namely
\begin{equation}
  \Phi = \Gamma x_2  + \phi \,.
\end{equation}
The scalar fluctuations $\phi$ are statistically homogeneous and obey
the equation
\begin{equation}
  \label{eq:phi3-1}
  \pp{\phi}{t} + u_i \pp{\phi}{x_i} = D \pp{^2\phi}{x_i^2} - \Gamma u_2\,,
\end{equation}
where $t$ is the time, $x_i$ the spatial coordinates, and $u_i$ denote
the velocity field.

The three-dimensional incompressible Navier-Stokes equations in the
vorticity formulation are solved together with Eq.~\eqref{eq:phi3-1}
by a dealiased pseudo-spectral approach,
cf.~\citet{canuto1988spectral}. Temporal integration is carried out by
a second order semi-implicit Adams-Bashforth/Crank-Nicolson
method. The integration domain is a triply periodic cube with length
$2\pi$. An external stochastic forcing, see
\citet{eswaran1988examination}, is applied to the velocity field to
maintain a statistically steady state. The forcing is statistically
isotropic and limited to low wave-numbers so that the small scales are
not affected by the forcing scheme. The simulations have been carried
out with an in-house hybrid MPI/OpenMP parallelized simulation code on
the supercomputer JUQUEEN at research center J{\"u}lich, Germany.

Characteristic parameters of the DNS are presented in
table~\ref{tab:R}, where $N$ denotes the number of grid points along
one coordinate axis, $\mathit{Re}_\lambda$ the Taylor based Reynolds
number, $\avg{k}$ the mean kinetic energy, $\avg{\varepsilon}$ the
mean energy dissipation, $\avg{\phi^2}$ the mean scalar variance,
$\avg{\chi}$ the mean scalar dissipation, and $\avg{u_2\phi}$ the mean
scalar flux. The production of scalar variance is
$-2 \avg{u_2 \phi} \Gamma$. Ensemble-averages are denoted by angular
brackets and are computed over the whole computational domain due to
homogeneity and over a time frame $t_{\rm avg}$ due to stationarity.
The number of analyzed ensembles is in the range between $M=6$ for
case R5 up to $M=189$ for case R0. Resolving the smallest scales by
the numerical grid is important for the accuracy of the DNS. To ensure
an appropriate resolution of the smallest scales,  the number of grid
points has been increased to as high as $4096 \times 4096 \times 4096$
for case R5. Following \citet{ishihara2007} and \citet{donzis2005}, a
resolution condition of $\kappa_{\rm max} \eta> 2.5$ is maintained for
all cases to accurately compute high-order statistics. Here,
$\kappa_{\rm max}$ denotes the highest resolved wavenumber and
$\eta=\nu^{3/4}\avg{\varepsilon}^{-1/4}$ denotes the Kolmogorov length
scale. Further details about the DNS are presented by
\citet{gauding2015line} and \citet{peters2016higher}.

\begin{table}
  \caption{\label{tab:R} Summary of different DNS cases. Reynolds number variation between
    ${\rm Re}_\lambda=88$ and ${\rm Re}_\lambda=529$.}
  \centering
    \begin{tabular}{l c c c c c c}\toprule
      &  R0      & R1 & R2 & R3 & R4 & R5  \\
      \midrule
      $ N$              & $512^3$    & $1024^3$    & $1024^3$    & $2048^3$    & $2048^3$    & $4096^3$\\
      $ {\rm Re}_\lambda$   & 88        & 119   & 184  & 215 & 331   & 529\\
      $\nu$                & 0.01    & 0.0055 & 0.0025 & 0.0019 & 0.0010 & 0.00048  \\
      $\avg{k}$            & 11.15   & 11.20   & 11.42    & 12.70   & 14.35   & 23.95 \\
      $\avg{\varepsilon}$  & 10.78  & 10.52   & 10.30    & 11.87   & 12.55   & 28.51 \\
      $\avg{\phi^{2}}$  & 1.95   & 1.89    & 1.94     & 2.47    &  2.25   &  2.41 \\
      $\avg{\chi}$         & 3.92  & 3.90    & 4.01     & 5.00    & 4.76    &  6.78 \\
      $-2 \Gamma\avg{ u_2\phi}$ & 3.93    & 3.98    & 4.03     & 4.95    & 4.79    & 5.76 \\
      $t_{\rm avg}/\tau$       & 100 & 30 & 30 & 10 & 10 & 2 \\
      $M$       & 189 & 62 & 61 & 10 & 10 & 6 \\
            $\kappa_{\rm max} \eta$  & 3.93      & 4.99    & 2.93    &  4.41    & 2.53
      & 2.95 \\
      \bottomrule
    \end{tabular}
\end{table}

\section{Scale-by-scale transport equations for even order structure
  functions} \label{section_theory} Starting from
Eq.~\eqref{eq:phi3-1}, a transport equation for the even moments of
the scalar increment can be derived by using a similar procedure as
presented in \cite{danaila1999YaglomForcing} and \citet{hill2001}. For
homogeneous turbulence this equation reads
\begin{equation}
  \label{eq:two-point2n-a}
  \pp{}{t}\big<(\Delta\phi)^{2n}\big>(\vec r) +
  \underbrace{\pp{}{r_i}\big< (\Delta
    u_i) (\Delta \phi)^{2n} \big> (\vec r)}_{-\mathit{Tr}_{2n}}
  + \underbrace{2 n \Gamma \big< (\Delta u_2) (\Delta \phi)^{2n-1}  \big> (\vec r)}_{-\mathit{Pr}_{2n}} 
  = J_{2n}(\vec r)  \,,
\end{equation}
where $J_{2n}$ are the molecular-diffusion terms, i.e.\
\begin{equation}
  \label{eq:j2n}
  J_{2n}(\vec r) = n D 
  \left\langle{
    (\Delta\phi)^{n-1}  
    \left[ 
      \pp{^2 (\Delta\phi)}{x_i'^2} +
      \pp{^2 (\Delta\phi)}{x_i^2}
    \right]
  } \right\rangle\,.
\end{equation}
The term $J_{2n}$ is a function of $\vec r$ and remains finite even
when $D$ tends to zero. Equation \eqref{eq:two-point2n-a} is exact,
which means that it is derived from the governing equations without
any approximation beside of incompressibility and homogeneity.  It is
convenient to decompose $J_{2n}(\vec r)$ by partial integration, i.e.\
\begin{equation}
  \label{eq:j2n-b}
  J_{2n}(\vec r) = \underbrace{2D \pp{^2}{r_i^2} \avg{(\Delta\phi)^{2n}}}_{D_{2n}} - \underbrace{n(2n-1) 
    \avg{(\Delta\phi)^{2n-2} \left( \chi(\vec x) + \chi(\vec x + \vec r)  \right)}}_{E_{2n}} \,.
\end{equation}

The terms of Eqs.~\eqref{eq:two-point2n-a} and \eqref{eq:j2n-b} can be
physically interpreted.  The first term on the left-hand side is the
temporal change of the moments of the scalar increment. The second
term $\mathit{Tr}_{2n}$ is a mixed velocity-scalar structure function
and represents an inter-scale transport from large scale towards small
scales. Therefore, it is hereinafter referred to as transport
term. The third term $\mathit{Pr}_{2n}$ is proportional to the mean
scalar gradient $\Gamma$ and can be interpreted as a production term
that is mainly active at large scales. According to
Eq.~\eqref{eq:j2n-b},  term $J_{2n}$ splits up into two terms. The
first is a diffusive transport term $D_{2n}$. The latter term is the
so-called dissipation source term $E_{2n}$.  The second order
dissipation source term $E_{2}$ is independent of $r$ and simplifies
to $2\avg{\chi}$. In this case, Eq.~\eqref{eq:two-point2n-a} reduces
to the Yaglom equation for homogeneous, but anisotropic turbulence at
finite Reynolds number, cf.~\citet{Antonia1997a}. For higher orders,
$E_{2n}$ is a multi-scale correlation between the local scalar
difference $(\Delta\phi)^{2n-2}$ and the sum of the instantaneous
scalar dissipation $\chi$ at two points separated by $\vec r$.

Most theoretical works consider integrated formulations of
Eq.~\eqref{eq:two-point2n-a}, where an integration is carried out
either over spheres of radius $r$ (when the data used comes from DNS,
\citet{Casciola_etal_JFM03}), or along straight lines with a length
$r$. This latter possibility is preferred by studies based on
experimental data, because of the use of Taylor's hypothesis.  In this
work we follow an approach introduced by \citet{peters2016higher} and
evaluate the terms of the scale-by-scale budget equations as given by
Eq.~\eqref{eq:two-point2n-a}. The
transport and the diffusive terms involve derivatives in
$\vec r$-space. By using the rules of differentiation the
$r_i$-derivative can be replaced by local spatial derivatives. For
example, with homogeneity and incompressibility, the transport term is
written as
\begin{equation}
  \label{eq:trans}
  \pp{}{r_i}\big< (\Delta u_i) (\Delta \phi)^{2n} \big>
  =  2n \left\langle{(\Delta u_i)(\Delta \phi)^{2n-1} \pp{\phi}{x_i}}\right\rangle \,.
\end{equation}
By using Eq.~\eqref{eq:trans},  the transport term can be easily
computed from DNS data, as only a correlation between velocity and
scalar increments with a local scalar derivative is required. The
diffusive term can be computed in the same way.
The temporal decay term is negligible, because the continuous forcing ensures stationary state of the mixing. 

\begin{figure}
  \centering
  \includegraphics[width=0.49\linewidth]{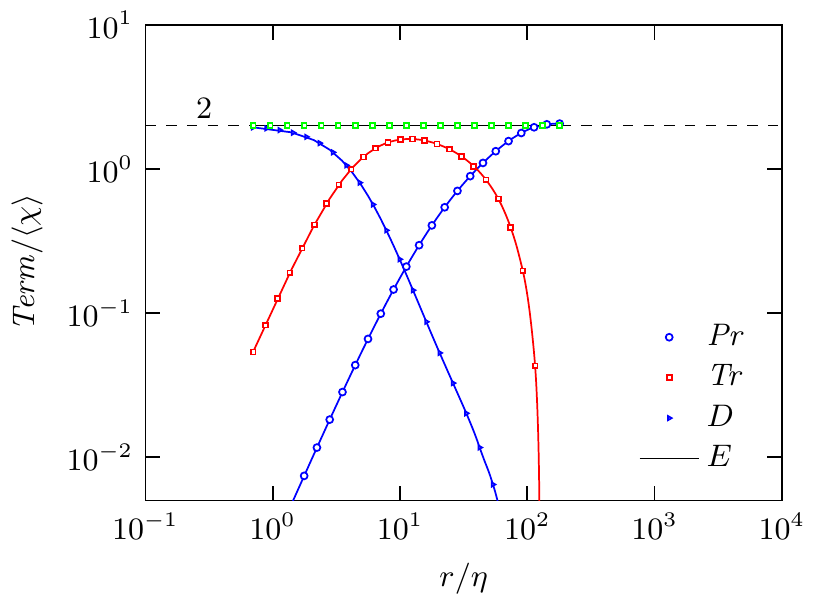}
  \includegraphics[width=0.49\linewidth]{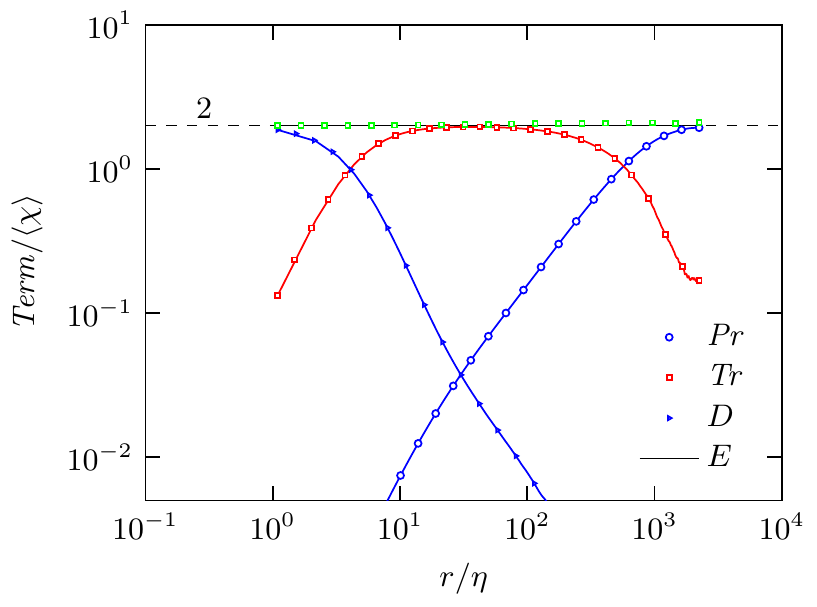}\\[2ex]
  \includegraphics[width=0.49\linewidth]{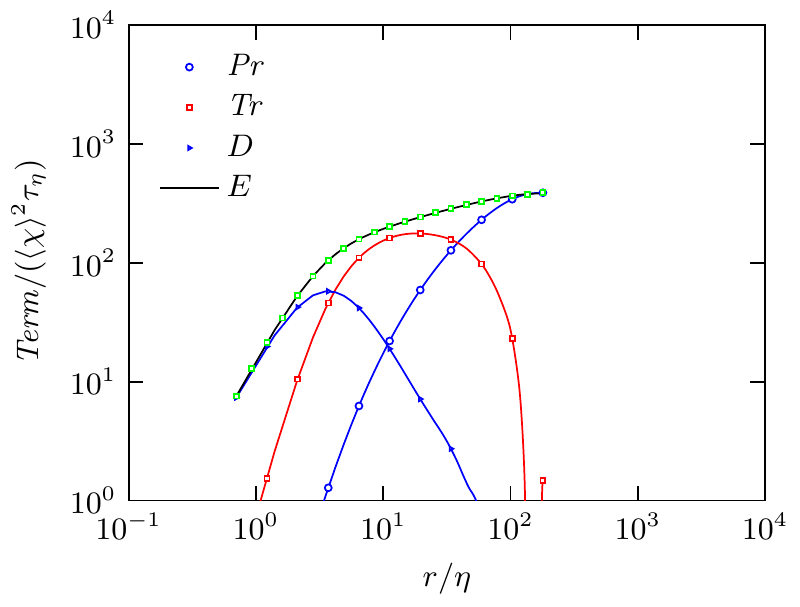}
  \includegraphics[width=0.49\linewidth]{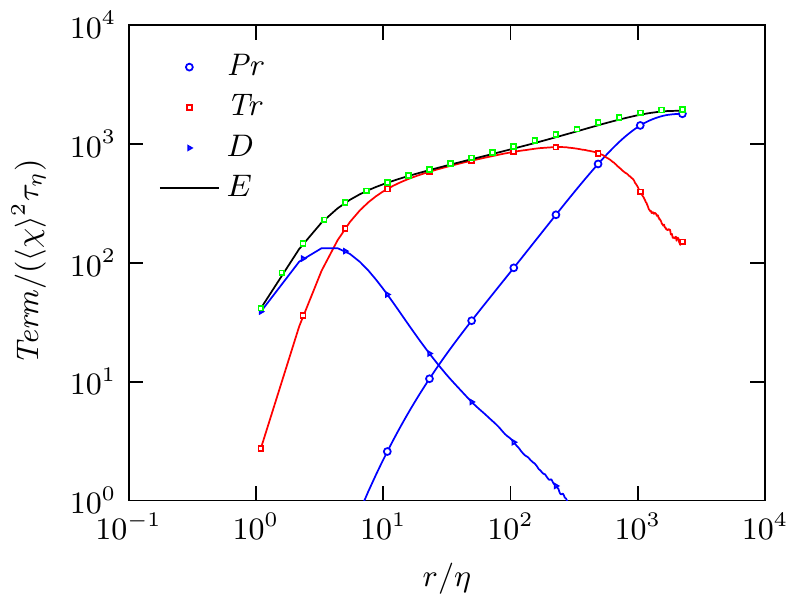}
  \caption{Terms in Eq.~\eqref{eq:two-point2n-a} and \eqref{eq:j2n-b}
    for the second (top) and fourth order scalar structure function
    equation. The left column is for case R0 and the right column is
    for case R5. The green squares represent the sum
    $\mathit{Tr}_{2n}(r) + \mathit{Pr}_{2n}(r) + D_{2n}(r)$, which
    balances the term $E_{2n}(r)$ for all scales. This indicates that
    the budget is satisfied. It is notable, that $E_2$ is independent
    of $r$ and simplifies to $2\avg{\chi}$.}
\label{fig:budget}
\end{figure}

Even though Eqs.~\eqref{eq:two-point2n-a} and \eqref{eq:j2n-b} are
exact, they cannot be solved directly, because they are not
closed. Nonetheless,  all terms can be computed from DNS. This is displayed in
fig.~\ref{fig:budget}, where the terms of the transport equations for
the second and fourth order scalar structure functions
$\avg{(\Delta \phi)^{2n}}$ are shown for $\mathit{Re}_\lambda=88$ and
$\mathit{Re}_\lambda=529$. % Figure \ref{fig:sf} provides an
% information about the scale dependence of the production, transport
% and dissipation mechanisms present in turbulent flows.
Let us first briefly discuss the scale-by-scale budget for the second
order scalar structure function.  For the second order the sum
$\mathit{Tr}_2 + \mathit{Pr}_2 + D_{2}$ is independent of $r$ and
equals $2\avg{\chi}$. The diffusive terms $D_2$ is dominant in the
dissipative range and tends to $2\avg{\chi}$ for $r\to 0$. The
production term $\mathit{Pr}_2$ in dominant at large scales and equals
$2\avg{\chi}$ for $r\to\infty$. The transport term is dominant in the
inertial range, but attains the value $2 \avg{\chi}$ only when the
Reynolds number is large enough.  For low Reynolds numbers, statistics
in the inertial range are strongly affected by molecular diffusion and
large scale effects originating from the mean scalar gradient. The
width of the inertial range scaling regime increases only slowly with
Reynolds number, which underlines that accounting for finite Reynolds
number effects is important. 

The scale-by-scale budget for the fourth order structure function
$\avg{(\Delta\phi)^4}$ is significantly different. Here, the
dissipative source term~$E_4$ is a function of the separation
distance, revealing two different scaling regimes for the dissipative
and the inertial range. In the dissipative range $E_4(r)$ scales with
$r^2$ and balances the diffusive term $D_4$. {\color{black} Term $E_4$
  is proportional to
  $\avg{(\Delta\phi)^2 \left( \chi(\vec x) + \chi(\vec x + \vec r)
    \right)}$.  Along any arbitrary direction $r_k$, a Taylor series
  development when $r_k \rightarrow 0$ for $(\Delta\phi)^2$ gives {\color{black}
  $(\Delta\phi)^2 = \left (\pp{\phi}{x_k} \right)^2 r^2$.} Injecting
  this in the expression of $E_4$, and because the average applies to
  the product between $(\Delta\phi)^2$ and $\chi$, the small-scale
  limit of $E_4$ will be proportional to the average value of the
  square of the scalar dissipation.  Finally, $E_4$} can be written in
the limit $r_k \to 0$ as
\begin{equation}
  \label{eq:e4_dr}
  E_4(r_k) = \frac{2 \avg{\chi^2}}{D} \,r_k^2 \,.
\end{equation}
{\color{black}The derivation of eq.~\eqref{eq:e4_dr} is detailed in
\ref{lab:der}.} In the inertial range, fig.~\ref{fig:budget} indicates
that $E_4$ is balanced by the transport term $\mathit{Tr}_4$, provided
that the Reynolds number is high enough. Under this condition, $E_4$
and $\mathit{Tr}_4$ have the same inertial range scaling exponent
$\zeta_4$ and the relation
\begin{equation}
  E_4(r) = \mathit{Tr}_4(r) = c_4 r^{\zeta_4}
\end{equation}
is satisfied. Note that due to internal intermittency, neither $c_4$
nor $\zeta_4$ can be obtained from dimensional arguments,
cf.~\cite{Kolmogorov1962}.   In the large scale limit, $E_4$ balances
$\mathit{Pr}_4$ and becomes independent of $r$, as it tends to
$12\avg{\phi^2\chi}$ for $r\to\infty$. These limits will be exploited
in Section ~\ref{subsection_sim_4thorder} to develop similarity scales for
higher-order moments.

%\section{Modeling of the transport term}

\section{Similarity scales pertaining to the dissipative range } \label{sec:similarity}
\begin{figure}
  \centering
  \includegraphics[width=0.49\linewidth]{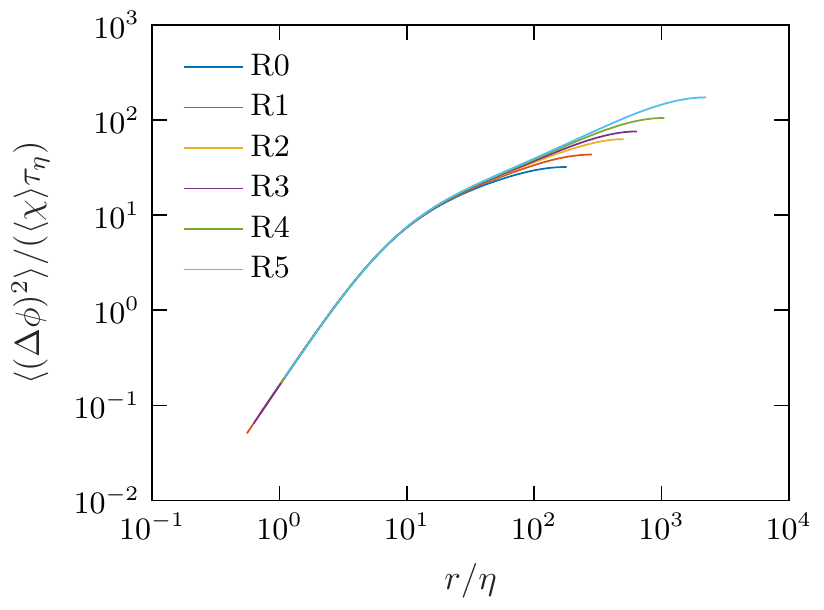}
  \includegraphics[width=0.49\linewidth]{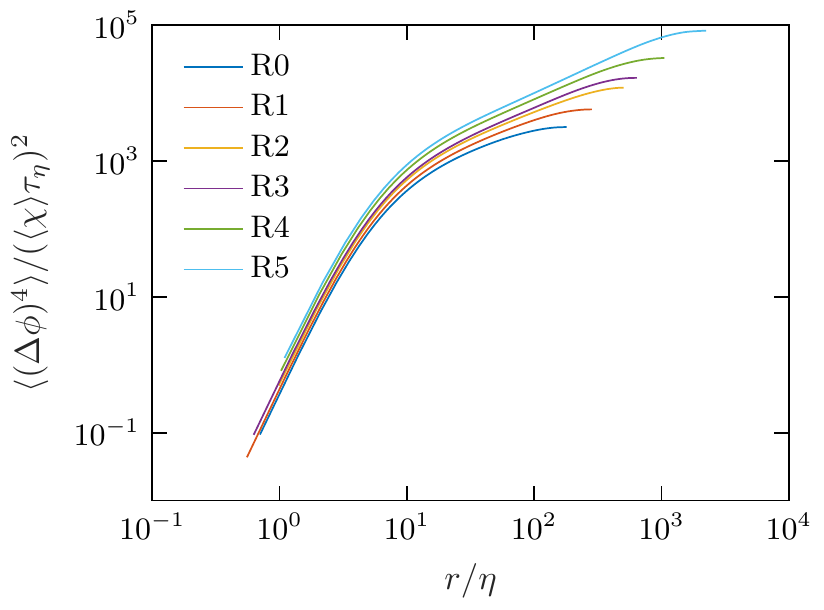}
  \caption{Second order (left) and fourth order (right) scalar
    structure function $\avg{(\Delta\phi)^{2n}}$ for all cases. The
    structure functions are normalized by $\avg{\chi}^n\tau_\eta^n$,
    with $\tau_\eta$ being the Kolmogorov time-scale, defined as
    $\tau_\eta=\nu^{1/2}\avg{\varepsilon}^{-1/2}$.}
  \label{fig:sf_ps_n24}
\end{figure}

\begin{figure}
  \centering
  \includegraphics[width=0.49\linewidth]{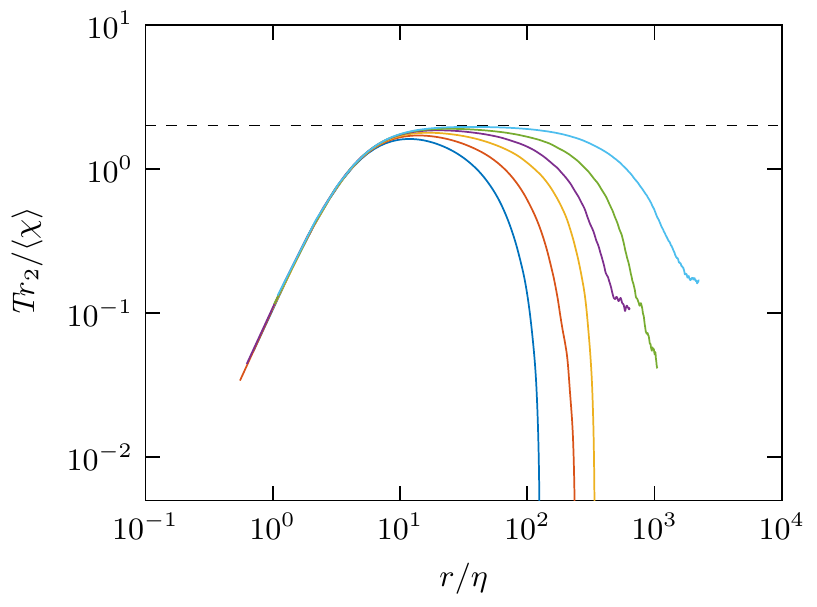}
  \includegraphics[width=0.48\linewidth]{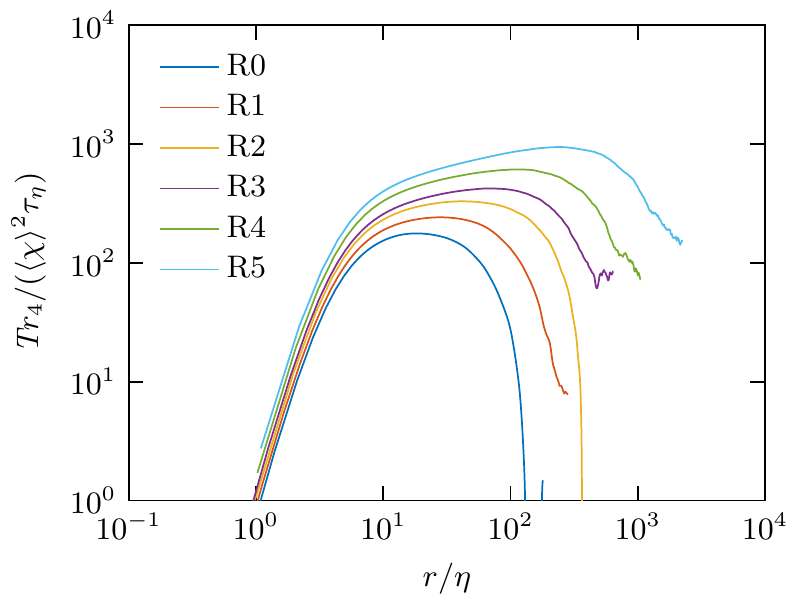}
  \caption{Second order (left) and fourth order (right) transport term
    $\mathit{Tr}_{2n}$ of Eq.~\eqref{eq:two-point2n-a} for all
    cases.}
  \label{fig:budget_n1}
\end{figure}

Kolmogorov-Obukhov-Corrsin (KOC) scales are built from the mean value
of the scalar dissipation rate $\left\langle \chi \right\rangle $.  As
illustrated in figures \ref{fig:sf_ps_n24} and \ref{fig:budget_n1},  an
adequate collapse after normalization using KOC scales only applies
for the second-order moments, and for the third-order mixed
velocity-scalar structure functions
$\avg{(\Delta u_i)(\Delta \phi)^2}$ when the Reynolds number is large
enough. Figures~\ref{fig:sf_ps_n24} and \ref{fig:budget_n1} clearly
emphasize that the mean dissipation $\avg{\chi}$ is not consistent
with a small-scale collapse of the fourth-order structure function
$S_4$,  or transport term $\mathit{Tr}_4$.  In the following, we use the
transport equations derived in Section~\ref{section_theory} to provide
expressions for the similarity scales.

\subsection{Similarity scales for the second-order moments transport equation} \label{subsection_sim_2ndorder}
 In the transport equation for the second-order moments, all terms depend on the spatial increment $\vec r$.  Therefore, anisotropy of the flow and mixing is accounted for.  
In the following, we address the question of self-similarity, as introduced by ~\cite{Townsend1976} for the one-point energy budget equation, and later on by ~\cite{George1992} for the two-point statistics in the spectral space, and  ~\cite{Antonia2003} for two-point statistics in real space.

In order to examine the conditions under which
Eq.~(\ref{eq:two-point2n-a}) satisfies similarity, we need to assume
functional forms for the terms in this equation. This equation, for
the second-order moments, can be formally written as:
$\mathit{Tr}_2+\mathit{Pr}_2+ D_2=(\avg{\chi(\vec x + \vec
  r)}+\avg{\chi(\vec x)})$, where $\mathit{Tr}_2$ is the transport
term, $\mathit{Pr}_2$ is the production, and $D_2$ is the molecular
term.  Because anisotropy is accounted for, and as the flow and mixing
are stationary and spatially homogeneous, these terms depend on the
vector $\vec r$ and vary as a function of the energy injected in the
flow, so say, on the Reynolds number of the flow. We thereby have to
distinguish between functions which are Reynolds-dependent (and do not
depend on the scale $\vec r$) and the functions which depend on the
spatial increment $\vec r$. Following ~\cite{Antonia2003} and
~\cite{Burattini2005a}, we take
\begin{eqnarray}
\avg{(\Delta \phi)^2}=c_1(Re) \cdot  f(\vec \xi) \nonumber \\
\frac{\partial}{\partial r_j}\avg{\Delta u_j (\Delta \phi)^2}=c_2(Re) \cdot g(\vec \xi), \nonumber \\
\Gamma \avg{ \Delta u_2 \Delta \phi}=c_3(Re) \cdot  v(\vec \xi), \nonumber \\   
\avg \chi^+ +\avg \chi^-= c_4(Re) \cdot  w(\vec \xi).  \label{sim_aniso_1}
\end{eqnarray}
where $\vec \xi= \vec r/{\cal L}$ and $\cal L$ is a characteristic length scale, to be determined. The scale $\cal L$ depends on $Re$, but also on the spatial direction of the separation vector 
$\vec r$, say ${\vec e}_r= \vec r/ |\vec r|$.  It is natural to identify $c_1\sim \phi_{ref}^2$, where the index 'ref' stands for the 'reference'. Similarly, 
$c_2\sim v_{ref} \phi_{ref}^2/\cal L$,  $c_3\sim \Gamma v_{ref} \phi_{ref}$  and $c_4\sim \chi_{ref}$.
 A dependence on the initial conditions, as explained by ~\cite{George1992} is also plausible.  The separation between functions of $Re$ and $\vec \xi$ allows solutions of the transport equation for which a relative balance among all of the terms is maintained for increasing Reynolds numbers.  Upon substituting Eqs.~(\ref{sim_aniso_1}) into Eq.~(\ref{eq:two-point2n-a}), considering the viscous term $D \frac{\partial^2}{\partial r_j^2} \avg{ (\Delta \phi)^2}$   as a reference, whose scaling is $D \frac{\phi_r^2}{{\cal L}^2}$,  we obtain 
\begin{equation}
\frac{v_{ref} \cdot {\cal L}}{D}= K_1;  \label{eq_ratio_K1}
\end{equation}
\begin{equation}
\frac{\avg {\chi}_{ref} \cdot {\cal L}^2}{D \cdot \phi_r^2}= K_2; \label{eq_ratio_K2}
\end{equation}
\begin{equation}
\frac{D \cdot \phi_{ref}}{v_{ref} \Gamma {\cal L}^2} = K_3. \label{eq_ratio_K3}
\end{equation}
The constants $K_i$ are understood as such with respect to the
variable $Re$. Combining these equations and considering that the
similarity scale for the velocity field is, at least for the smallest
scales, proportional to the Kolmogorov scale
$\eta=(\nu^3/\avg \varepsilon)^{1/4}$, then a characteristic scale of the
scalar can be identified with
$\phi_{ref} \equiv [\mathit{Sc}\, \tau_\eta \avg \chi]^{1/2}$, where $\tau_\eta$ is the
Kolmogorov time scale. Thus, the energy transport term $Tr_2$ should
scale as
$Tr_2 \sim v_{ref} \frac{\phi_{ref}^2}{\cal L}=\eta \frac{\mathit{Sc}\, \tau_K \avg
  \chi}{\eta}=\mathit{Sc} \cdot \avg \chi $. For our simulations, the Schmidt
number is equal to $1$, so term $Tr_2$ should scale as $\avg \chi $
when the energy injected in the flow varies. In other words, the ratio
$Tr_2/ \avg \chi $ may vary as a function of the vectorial separation
$\vec \xi$, but must be a constant of the Reynolds number, if the
similarity was to be valid.  This is firmly confirmed by the excellent
collapse of the normalized curves in Fig. \ref{fig:budget_n1} for the
smallest scales. For increasing Reynolds numbers, the agreement is
excellent over a wider and wider range of scales, thus validating our
approach.
 
%the Taylor microscale $\lambda \equiv \sqrt{\frac{Q \overline{\nu}}{\overline{\epsilon}_{VV}}}$, therefore ${\cal L} \equiv \lambda$.  Similarly to what was formally found for grid turbulence or for shear flows~\cite{George1992,Antonia2003,GeorgeGibson1992,Sadeghi2015} the Taylor microscale emerges as one relevant length scale for the whole range of scales. 
We may preclude that in flows where  similarity is tenable at the smallest scales only, then the unique solution of the problem that emerges is the Kolmogorov velocity and length scales, and KOC scale for the scalar itself.

\subsection{Similarity scales for the higher-order moments transport equations} \label{subsection_sim_4thorder}
 
\begin{figure}
  \centering
  \includegraphics[width=0.49\linewidth]{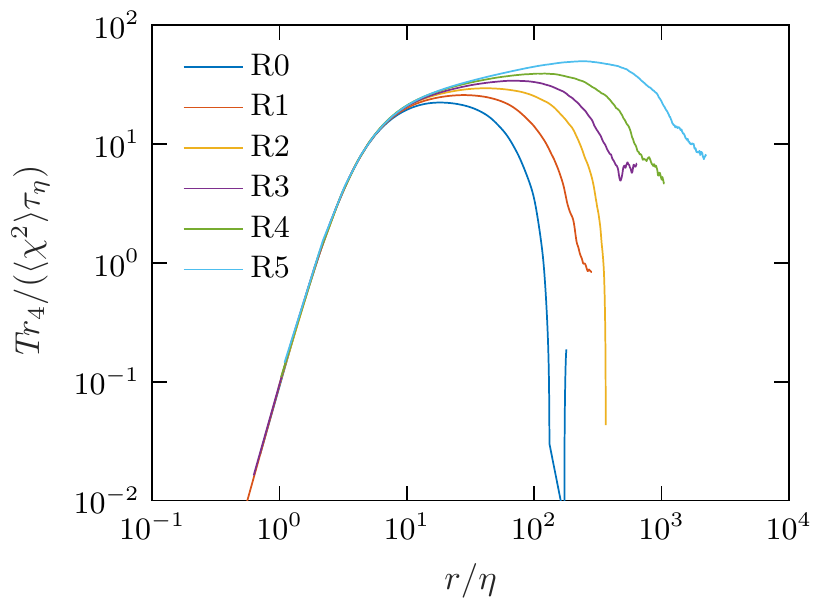}
  \caption{Fourth order transport term $\mathit{Tr}_4$ normalized by
    similarity scales $\avg{\chi^2} \tau_\eta$ from
    eq.~\eqref{eq_scaling_Tr_dr_2}.}
  \label{fig:Trn}
\end{figure}

A similar analysis may be performed for the transport equation of the
high-order moments. The development is done here for the 4-th order
moments, but the generalization is then straightforward for arbitrary
higher-order moments.  The transport term $\mathit{Tr}_4$, scales as
$\mathit{Tr}_4 \sim v_{ref} \frac{\phi_{ref}^4}{\cal L}$. An important point
to be underlined here is that, similarity scales for high-order
moments are not the same as those for the second-order and third-order
moments. For the sake of simplicity, we keep however the notation
$v_{ref}$ and $\cal L$.  The molecular term scales as
$D_4 \sim D \frac{\phi_{ref}^4}{{\cal L}^2}$, the production term behaves
as $Pr_4 \sim v_{ref} \Gamma \phi_{ref}^3$. Lastly, the dissipation source
term scales as $E_4 \sim \langle{\phi^2} \cdot \chi\rangle_{ref}$.  Note
that the latter term was not decomposed in a product of two terms, and
further reliable closures will be required to progress towards clear
scaling laws within the inertial range.  For the similarity to be
valid, ratios between any two terms should be real constant (i.e., not
functions of the Reynolds number, nor the scale). Therefore, several
results may be deduced, such as
$\mathit{Tr}_4/\mathit{Pr}_4={\rm const.}$, which translates in
\begin{equation}
\phi_{ref}\sim \Gamma \cal L, 
\end{equation}
or, the ratio $Tr_4/D_4={\rm const.}$ which is perfectly consistent
with Eq. (\ref{eq_ratio_K1}). Further information on the similarity
scale can be inferred for the dissipative range, so when
$\vec r \rightarrow 0$. For these scales,
\begin{equation}
E_4 \sim \frac{\langle \chi^2 \rangle {\cal L}^2}{D},  \label{eq_scaling_Edr}
\end{equation}
which should be proportional to the molecular term, as these two terms
are dominant in the dissipative range and they balance each other,
cf.\ Eq.~\eqref{eq:e4_dr}.  Therefore,
\begin{equation}
D \frac{\phi_{ref}^4}{{\cal L}^2} \sim \frac{\langle \chi^2 \rangle {\cal L}^2}{D},  \label{eq_scaling_balance_dr}
\end{equation}
which leads to
\begin{equation}
{\phi_{ref}^4} \sim \frac{\langle \chi^2 \rangle {\cal L}^4}{D^2}.  \label{eq_scaling_phi_dr}
\end{equation}
Note the clear dependence on $\langle \chi^2 \rangle$, which is different from the square of $\langle \chi\rangle$. 
Furthermore, scalings for all the other terms in the dissipative range may be inferred, such as, for example, the transport term $\mathit{Tr}_4$, which writes 
\begin{equation}
\mathit{Tr}_4  \sim \frac{v_{ref} {\phi_{ref}^4}}{\cal L}.  \label{eq_scaling_Tr_dr}
\end{equation}
After some straightforward manipulations, and {\bf if the similarity
  scale is the same as that of the velocity field}, so $\cal L=\eta$, then 
\begin{equation}
\mathit{Tr}_4  \sim \tau_\eta Sc^2 \langle \chi^2 \rangle.  \label{eq_scaling_Tr_dr_2}
\end{equation}
%where $\tau_\eta$ is the Kolmogorov time.  
This scaling is reasonably supported by fig.  \ref{fig:Trn}, at least for the smallest
scales, and for the value $Sc=1$ of our simulations.  The non-perfect collapse
is most likely due to the fact that the best adapted similarity length
scale, $\cal L$, should be built from higher-order moments of the
energy dissipation rate, so from $\langle \varepsilon^2 \rangle$, and not
from $\langle \varepsilon\rangle$ itself. This hypothesis is being
explored and is subject of future publication.

\begin{figure}
  \centering
  \includegraphics[width=0.49\linewidth]{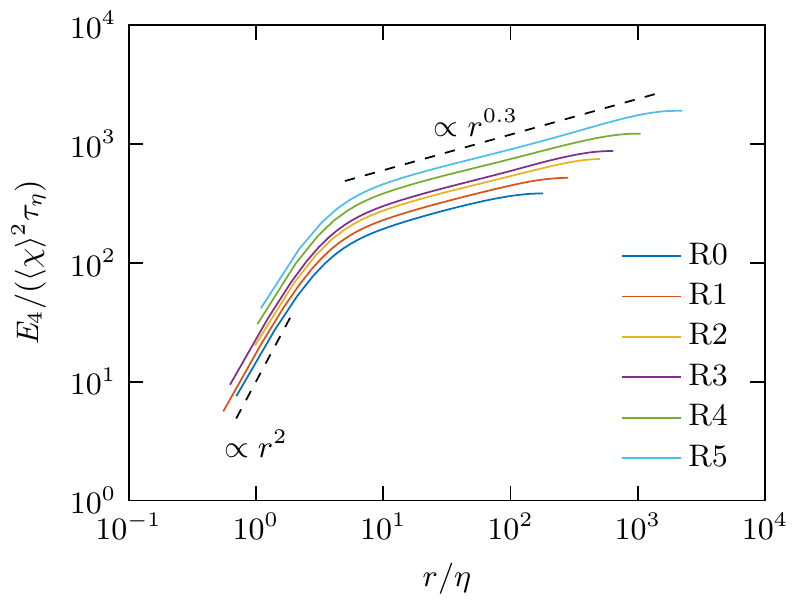}
  \includegraphics[width=0.49\linewidth]{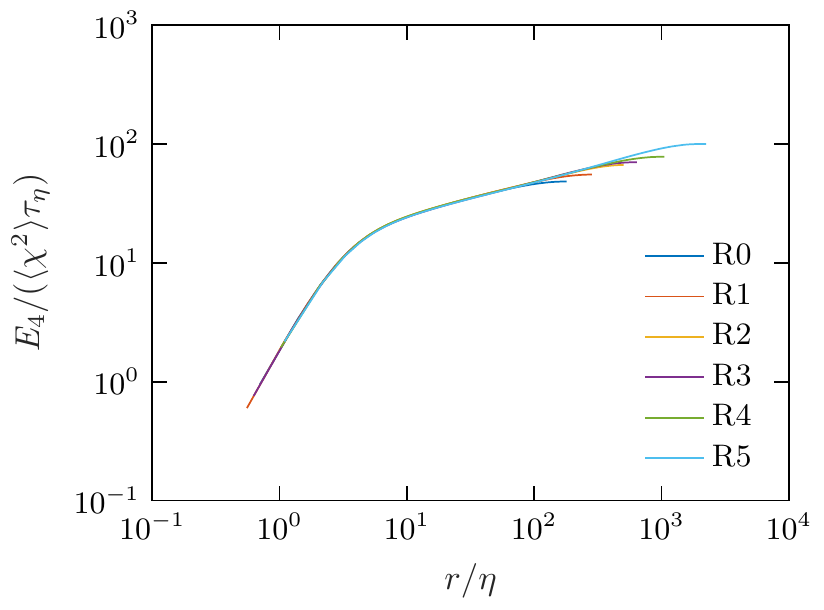}
  \caption{Fourth order dissipation source term $E_4$ normalized by
    the KOC scaling variables $\avg{\chi}^2\tau_\eta$ (left) and the
    higher-order similarity scales $\avg{\chi^2}\tau_\eta$ (right)}
  \label{fig:E}
\end{figure}

The scaling of the dissipation source term $E_4$ is of special
interest, because this term depends on the scale $r$ and it balances one of the
other terms of Eq.~\eqref{eq:two-point2n-a}. Figure~\ref{fig:E} shows
the term $E_{4}$ for the different Reynolds numbers under
consideration, where $E_4$ is normalized by the classical KOC
quantities (left) and by the similarity scales for the fourth order
moment (right) obtained from Eq.~\eqref{eq_scaling_Edr}. The
normalization by the KOC quantities leads to a staggered agreement
where the normalized $E_{4}(r)$ clearly depends on the Reynolds
number. This behavior was already expected by \citet{Landau1959}, who
argued that $\avg{\varepsilon}$ (and $\avg{\chi}$ for the scalar)
could not be the relevant normalization quantity for higher
orders. However, the modified scaling from Eq.~\eqref{eq_scaling_Edr}
reveals an excellent collapse of the curves independently of Reynolds
number. It is notable that the collapse is not limited to the
dissipative range, but extends up to the inertial range.  Thus,
Eq.~\eqref{eq_scaling_Edr} predicts correctly the Reynolds number
dependence of the dissipation source term $E_4(r)$.  {\color{black}Note also that, for the range of investigated
Reynolds numbers, DNS reveals that $\zeta_4$ is close to 0.3, cf.\
fig.~\ref{fig:E}.}

\begin{figure}
  \centering
    \includegraphics[width=0.49\linewidth]{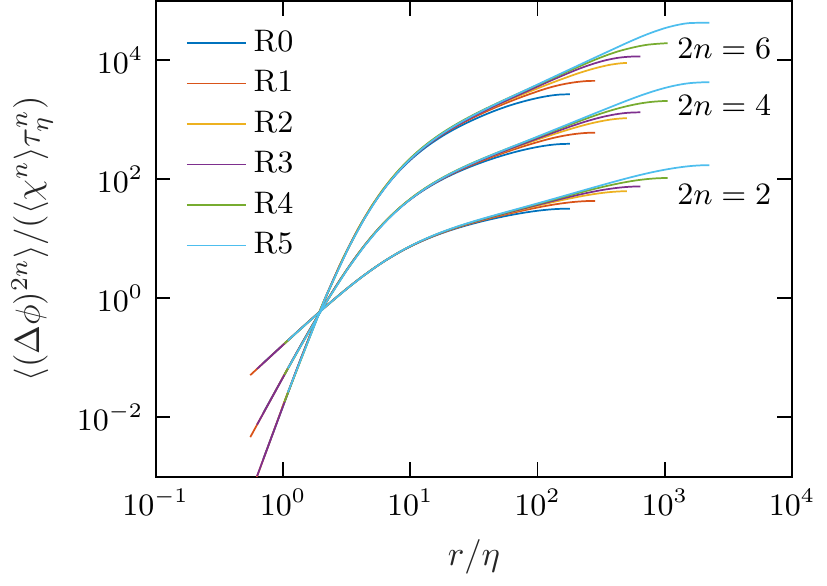}
    \caption{Higher order scalar structure functions
      $\avg{\Delta\phi}^{2n}$ normalized according to
      Eq.~\eqref{eq:scaling}.}
  \label{fig:Sn}
\end{figure}

As demonstrated before, the classical KOC scaling is not valid for
higher-order structure functions. Following the similarity scales for
higher-order moments, cf.\ Eq.~\eqref{eq_scaling_phi_dr}, a scaling
relation can be derived for scalar structure functions of any even
order
 \begin{equation}
  \label{eq:scaling}
  \frac{\avg{(\Delta\phi)^{2n}}}{\avg{\chi}^{n} \tau_\eta^{n}} = \,
C_{2n} \frac{\avg{\chi^{n}}}{\avg{\chi}^{n}}\, \left(
  \frac{r}{\eta} \right)^{2n} \,.
\end{equation}
The prefactor $C_{2n} \avg{\chi^{n}}/\avg{\chi}^{n}$ on the right-hand
side accounts for the dependence of the normalized $2n$-th order
structure function on the Reynolds number. The constant $C_{2n}$ is a
function of the order $n$, but not of the Reynolds number. For $n=1$,
eq.~\eqref{eq:scaling} reduces to the classical KOC scaling. The
scaling from Eq.~\eqref{eq:scaling} is supported by fig.~\ref{fig:Sn},
where the normalized structure functions $\avg{(\Delta\phi)^{2n}}$ are
shown for the second, fourth and sixth order. By using
Eq.~\eqref{eq:scaling} the quality of the collapse of the higher-order
structure functions is as good as the collapse for the second order,
and significantly improved compared to fig.~\ref{fig:sf_ps_n24}
(right). We want to emphasize that it is not possible to derive
Eq.~\eqref{eq:scaling} from pure dimensional arguments, because
$\avg{\chi}^2$ and $\avg{\chi^2}$ have the same dimensions.

%\section{Closure for the transport term}
%
%\begin{figure}
%	\centering
%	\includegraphics[width=0.49\linewidth]{R0_approx_T2.pdf} \hfill
%	\includegraphics[width=0.49\linewidth]{R5_approx_T2.pdf}\\[2ex]
%	\includegraphics[width=0.49\linewidth]{R0_approx_T4.pdf} \hfill
%	\includegraphics[width=0.49\linewidth]{R5_approx_T4.pdf}
%	\caption{Closure of the transport term with a gradient flux
%		approximation and a decorrelation assumption for case R0 (left)
%		and case R5 (right). Second-order (top) and fourth-order
%		(bottom).}
%		\label{fig:model}
%		\end{figure}
 
\newpage
  
\section{Conclusions}
\label{sec:conclusions}
Transport equations for even-order structure functions were written
for a passive scalar mixing fed by a mean scalar gradient, with a
Schmidt number $Sc=1$.  Direct numerical simulations (DNS), in a range
of Reynolds numbers $R_\lambda \in [88,529]$ were performed and used
to assess the validity of these equations, for the particular cases of
second-order and fourth-order moments.  The involved terms pertain to
molecular diffusion, transport, production, and dissipative-fluxes.
The latter term, present at all scales, was shown to be  equal to:  \\
i) the mean scalar variance dissipation rate,  $\langle \chi \rangle$,  for the second-order moments transport equation; \\
ii) non-linear correlations between the instantaneous $\chi$ and the second-order scalar increment, for the fourth-order moments transport equation. \\
The  equations were further analysed to show that the
si\-mi\-la\-ri\-ty scales (i.e., variables which allow for perfect collapse of
the normalised terms in the equations) are, for second-order moments,
fully consistent with KOC
theory. However, for higher-order moments, adequate similarity scales
are built from $\langle \chi^n \rangle$.  The similarity is tenable
for the dissipative range, and the beginning of the scaling
range (or, inertial range). 

Finding the similarity variables for larger scales requires
reliable closures of two other terms: the transport term ($Tr_{2n}$) ,
as well as of term $E_{2n}$.  The latter term is exactly closed only
for $n=1$, so for the second-order moments. For larger values of $n$,
the coupling between the instantaneous dissipation rate $\chi$
(specific of the small scales) and the large-scale moments
$(\Delta \phi)^{(2n-2)}$ reflects a direct correlation between large
and small scales, and it admits to an exact analytical development for
the small scales only. For larger scales, modelling is needed.

Another comment is devoted to the observation that, the passive scalar
being transported by the velocity field, characteristic scales of the
scalar and of the dynamic field, act together.  Our concern being on
mixing at $Sc=1$, our reasonable assumption was that the
characteristic scales were the same for both the scalar and dynamic
fields.  Nonetheless, as already noted, these scales depend on the
order $2n$ of the investigated moments, as clearly shown analytically
and proven by our DNS results.  This does not comply with the idea of
complete self-preservation (which requires a single
length/velocity/scalar scale,  \cite{Townsend1976}). Therefore, albeit
the value of the Schmidt number is $Sc=1$, and despite the fact
that the investigated Reynolds number is as high as $529$, this flow
and mixing are only incomplete self-preserving, i.e. only small scales
may be self-similar, when normalized with respect to quantities
adequately chosen.  Their  expressions are deduced basically from the
first principles, and in particular from transport equations of
high-order moments.
  
{\color{black} The physical signification of the second-oder moments is
  the energy at scales smaller or equal to $r$ (see also
  \cite{Danaila2012}). The physical parameters describing their
  evolution are the viscosity, the Prandtl (or Schmidt) number, and
  the mean value of the dissipation, in agreement with the classical
  KOC theory.  The fourth-order moments pertain to the variance of the
  variance of the signal, thus having a much more refined insight into
  the temporal activity of the flow. The theory shows that, at least
  for the smallest scales of the mixing, it is the mean value of the
  square of the dissipation which must be accounted for, corresponding
  to the dissipation of the variance of the variance at a given scale
  $r$.  At least for the range of Reynolds numbers investigated in
  this paper, the scale for four-order moments is clearly different
  from  the classical KOC scale. Whether or not these scales
  become equivalent in the limit of larger and larger Reynolds
  numbers, remains for now an open issue.}

 \vspace{2cm}
 
 We are honored to dedicate this paper to Prof. Andrew Pollard, for the celebration of his whole carreer, including impressive scientific contributions, editor work,  as well as the organisation  of many congresses.   L.D. acknowledges extended  fruitful discussions on quantifying internal intermittency from measurements with flying hot wires.    
 
 %\vspace{2cm}

 \section*{Acknowledgment}
 Financial support was provided by the Labex EMC3, under the grant VAVIDEN, as well as the Normandy Region and FEDER.   Additionally, the authors gratefully acknowledge the computing time
 granted on the supercomputer JUQUEEN at research center J{\"u}lich by
 the John von Neumann Institute for Computing
 (\cite{stephan2015juqueen}).

 \appendix
 {
\color{black}
\section{Exact relations between the moments of the scalar dissipation
and the scalar gradients}
\label{lab:der}
In this appendix we derive Eq.~\eqref{eq:e4_dr} under the assumptions
of local isotropy and homogeneity. With the Taylor series expansion
$(\Delta\phi)^2=\left( \pp{\phi}{x_1} \right)^2 r^2$ and
$\chi=2D \left(\pp{\phi}{x_i} \right)^2$, term $E_4(r)$ can be
written in the limit $r\to 0$ as
\begin{equation}
  \label{eq:E4a}
  E_4(r) = 24 D \left\langle{\left( \pp{\phi}{x_1} \right)^2
    \left[
      \left( \pp{\phi}{x_1} \right)^2 +
      \left( \pp{\phi}{x_2} \right)^2 +
      \left( \pp{\phi}{x_3} \right)^2 
    \right] }\right\rangle r^2 \,,
\end{equation}
where we assumed without loss of generality that the separation vector
$\vec r$ is aligned with the $x_1$~axis. The fourth-order derivatives
appearing in Eq.~\eqref{eq:E4a} can be directly related to
$\avg{\chi^2}$. Under the assumption of local isotropy the general
form of a fourth order gradient tensor reads (e.g. \cite{Siggia_PoF_81})
\begin{equation}
  \label{eq:tensor}
  \left\langle \pp{\phi}{x_i}  \pp{\phi}{x_j} \pp{\phi}{x_k} \pp{\phi}{x_l} \right\rangle
  =
  \alpha \delta_{ij} \delta_{kl} +
  \beta  \delta_{ik} \delta_{jl} +
  \gamma \delta_{il} \delta_{jk}  \,.
\end{equation}
The value of this generic tensor must be invariant under permutations
of its components which leads to 
\begin{equation}
  \left\langle \left( \pp{\phi}{x_1} \right)^4\right\rangle
  = 3 \left\langle \left( \pp{\phi}{x_1} \right)^2 \left( \pp{\phi}{x_2}
    \right)^ 2\right\rangle
  = 3 \left\langle{\left( \pp{\phi}{x_1} \right)^2 \left( \pp{\phi}{x_3}
    \right)^ 2} \right\rangle \,,
\end{equation}
and consequently
\begin{equation}
  \label{eq:E4b}
  \lim_{r \to 0} E_4(r) = 40 D \left\langle \left( \pp{\phi}{x_1} \right)^4 \right\rangle
  r^2 \,.
\end{equation}
Equation \eqref{eq:tensor} further implies that the second moment of
the scalar dissipation can be written as
\begin{equation}
  \label{eq:sd2}
  \avg{\chi^2} = 20 D^2 \left\langle \left( \pp{\phi}{x_1} \right)^4 \right\rangle \,,
\end{equation}
which gives with Eq.~\eqref{eq:E4b} the final result
\begin{equation}
  \lim_{r \to 0} E_4(r) = 2 \frac{\avg{\chi^2}}{D} r^2 \,.
\end{equation}

A generalization of Eq.~\eqref{eq:sd2} to higher order moments is
straightforward, and exact relations between the moments of the
scalar derivatives and the scalar dissipation are found, i.e.
\begin{equation}
  \label{eq:sdn}
  \left\langle \left( \pp{\phi}{x_1} \right)^{2n} \right\rangle = C_n
  \frac{\avg{\chi^n}}{D^n} \,,
\end{equation}
where the $C_n$ are Reynolds number independent coefficients. A
confirmation of Eq.~\eqref{eq:sdn} is shown in fig.~\ref{fig:sdn} for
a wide Reynolds numbers range and for $n$ between 1 and 4. Figure
\ref{fig:sdn} also indicates that the assumption of local isotropy is
well justified for the present DNS.

\begin{figure}
  \centering
    \includegraphics[width=0.49\linewidth]{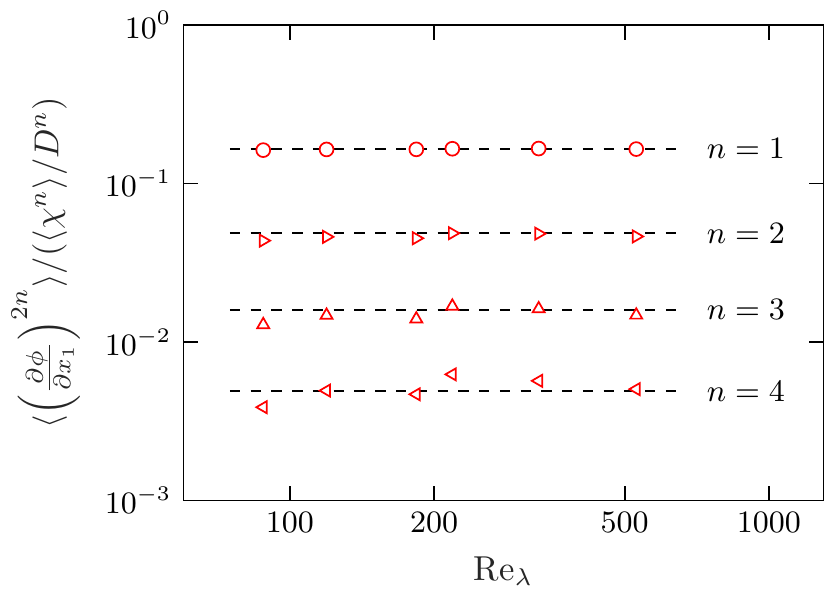}
    \caption{Scaling of the coefficient $C_n$ of eq.~\eqref{eq:sdn}
      for different orders and Reynolds numbers. }
  \label{fig:sdn}
\end{figure}
}


\newpage

\bibliographystyle{elsarticle-harv} 
%\bibliography{main}
\bibliography{main,biblio_Lea_large}

\end{document}